\documentclass[twocolumn]{aastex62}
\graphicspath{{./}{figures/}}
\usepackage{units}
\usepackage{multirow}

\shorttitle{Identification of supermassive binary black holes}
\shortauthors{Zhu \& Thrane}

\begin{document}

\title{Toward the unambiguous identification of supermassive binary black holes through Bayesian inference}
\correspondingauthor{Xing-Jiang Zhu}
\email{zhuxingjiang@gmail.com}

\author[0000-0001-7049-6468]{Xing-Jiang Zhu}
\affiliation{School of Physics and Astronomy, Monash University, Clayton, VIC 3800, Australia}
\affiliation{OzGrav: Australian Research Council Centre of Excellence for Gravitational Wave Discovery, Clayton, VIC 3800, Australia}

\author[0000-0002-4418-3895]{Eric Thrane}
\affiliation{School of Physics and Astronomy, Monash University, Clayton, VIC 3800, Australia}
\affiliation{OzGrav: Australian Research Council Centre of Excellence for Gravitational Wave Discovery, Clayton, VIC 3800, Australia}

\begin{abstract}
Supermassive binary black holes at sub-parsec orbital separations have yet to be discovered, with the possible exception of blazar OJ~287.
In parallel to the global hunt for nanohertz gravitational waves from supermassive binaries using pulsar timing arrays, there has been a growing sample of candidates reported from electromagnetic surveys, particularly searches for periodic variations in optical light curves of quasars.
However, the periodicity search is prone to false positives from quasar red noise and quasi-periodic oscillations from the accretion disc of a single supermassive black hole---especially when the data span fewer than a few signal cycles.
We present a Bayesian method for the detection of quasar (quasi-)periodicity in the presence of red noise.
We apply this method to the binary candidate PG1302$-$102, and show that a) there is very strong support (Bayes factor $>10^6$) for quasi-periodicity, and b) the data slightly favour a quasi-periodic oscillation over a sinusoidal signal, which we interpret as modest evidence against the binary black hole hypothesis.
We also find that the prevalent damped random walk red-noise model is disfavored with more than 99.9\% credibility.
Finally, we outline future work that may enable the unambiguous identification of supermassive binary black holes.
\end{abstract}


\section{Introduction} \label{sec:intro}
Supermassive binary black holes are thought to be common in the Universe as natural products of galaxy mergers \citep{Begelman80}.
They are likely to be the primary sources of nanohertz gravitational waves, which have been actively searched for using pulsar timing arrays in the past decade \citep[see, e.g.,][and references therein]{IPTA10,IPTA16,IPTA19}.
Current pulsar timing arrays are sensitive to individual binaries, with component black hole masses $\gtrsim 10^{9} M_{\odot}$ and binary orbital periods $\lesssim 10$ years (or orbital separations $\lesssim 0.01$ pc), up to distances of $\sim 200$ Mpc \citep{ZhuPPTACW14,EPTAcw15,NANOGrav19cw}, though see~\cite{Rosado} for the possibility of detecting high-redshift binaries.
No detection has been made so far, although it has been suggested that the detection of a stochastic gravitational-wave background formed by the combined emission from binaries across the Universe is likely in a few years \citep[e.g.,][]{AreWeThere,Cordes_Whitepaper19}.

The electromagnetic identification of sub-parsec supermassive binary black holes has proven to be challenging.
Direct imaging of such close binaries is only possible for sources within $\sim 100$ Mpc via very-long-baseline radio interferometry observations.
The closest binary with confident direct images, found in the radio galaxy 0402+379, has a separation of 7.3 pc \citep{Rodrigues06_0402}, corresponding to an orbital period of $\sim 10^4$ years \citep{Bansal17_0402+379}; see also \citet{Kharb17_subpcBBH} for a $\unit[0.35]{pc}$ binary candidate at $\unit[116]{Mpc}$ in the Seyfert galaxy NGC~7674.
How binaries like these can reach milli-parsec orbital separations, where the emission of gravitational waves can efficiently drive the binary to merge, is referred to as the ``final-parsec problem'', which remains an active area of theoretical investigations \citep[e.g.,][]{Milos01_finalpc,YuQJ02,Colpi14_review,Khan16_swiftmerge,Ryu18_triple,Sesana18_gas,LaiDong20_cbd,Taylor17,Chen19MN}.
Finding a sub-parsec supermassive binary black hole would not only provide insights into the final-parsec problem, but also shed light on the expected stochastic gravitational-wave signal strength for pulsar timing arrays \citep{Zhu19_GWBminmax,Mingarelli19_quasar}.

Periodically variable active galactic nuclei provide an interesting class of candidates for sub-parsec supermassive binary black holes.
The most prominent object is OJ 287, a blazar which exhibits $\unit[12]{year}$ quasi-periodic outbursts in optical light curves that date back to the 19th century \citep{OJ287_88,OJ287Nature08}.
This periodicity has been interpreted as the secondary black hole crossing the accretion disc of the primary in an eccentric orbit \citep{OJ287spin16,Dey19_oj287}.\footnote{After this paper was submitted, \cite{Laine} reported Spitzer observations of the predicted Eddington flare from OJ 287, adding significant weight to the theory that it is, in fact, a supermassive binary black hole.}
In the past five years, with the growth of time-domain astronomy, there have been a large number of sub-parsec supermassive binary black hole candidates claimed by various groups.

Based on the Catalina Real-time Transient Survey (CRTS), \citet{Graham15} put forward 111 binary black hole candidates from an optical variability analysis of 243,500 quasars.
Among them, PG1302$-$102 was the most significant candidate with a measured period of $\unit[1884]{days}$ \citep{PG1302Nature}.
This periodicity has been attributed to the relativistic Doppler boosting of emission from a mini-disc around the secondary black hole \citep{DOrazio15_boost}.
\citet{Charisi16PTF} reported 33 binary candidates from a periodicity search in a sample of 35,383 quasars in the photometric database of the Palomar Transient Factory.
\citet{LiuT15} identified a periodic variability of $\unit[542]{days}$ in quasar PSO J334.2+01.4 using data from the Pan-STARRS1 Medium Deep Survey. However, such a detection was found to be insignificant in a subsequent analysis of extended data \citep{LiuT16}; see \citet{Liu19} for a systematic search over 9000 quasars which resulted in one candidate in their extensive analysis.

Following the periodicity report of PG1302$-$102, \citet{Vaughan16_false} cautioned that the stochastic variability of normal quasars (i.e., those that do not host binary black holes) can resemble a periodic feature in light curves that span only a few periods and highlighted the importance of careful evaluation of false alarm rate.
Through Bayesian model comparison, \citet{Vaughan16_false} found a red-noise model is significantly favoured over a sinusoidal model for PG1302$-$102 based on the CRTS data.
More recently, \citet{LiuT18} revisited the periodicity of PG1302$-$102 using additional data from the All-Sky Automated Survey for Supernovae (ASAS-SN).
They employed a maximum likelihood method to search for a periodic signal in the presence of quasar red noise and found that the inclusion of ASAS-SN data reduced the periodicity significance. Therefore, \citet{LiuT18} concluded that the binary black hole model was disfavoured for PG1302$-$102.
\citet{Kova19_pg1302perturb} proposed a model that posits a cold spot in the accretion disc of the primary black hole. 
Such a model can produce a perturbed sinusoidal feature and thus explain the apparent decrease in periodicity significance found by \citet{LiuT18}.

Here, we propose a fully Bayesian framework for the identification of supermassive binary black hole candidates in time-domain electromagnetic surveys. It is capable of dealing with generic signal forms in the presence of red noise. Our work improves on previous studies in several ways. First, our method allows the inference of noise properties and signal parameters simultaneously and thus accounts for potential covariance between a periodic signal and quasar red noise. Second, it is robust to offsets in data collected with different surveys and possible over/under-estimation of measurement uncertainties. Third, we adopt a general form of quasar red noise and search for deviation from the commonly assumed damped random walk (DRW) model \citep{Kelly09DRW}.
We also compare the sinusoidal signal hypothesis against a quasi-periodic oscillation (QPO) model based on behavior observed in many stellar-mass black hole X-ray binaries \citep[see, e.g.,][for a recent review]{Motta16_QPO}.

The remainder of this paper is organized as follows.
In Section \ref{sec:method}, we describe the Bayesian inference framework and introduce our signal and noise models.
In Section \ref{sec:pg1302}, we apply the method to PG1302$-$102 with data from CRTS, ASAS-SN, and the Lincoln Near-Earth Asteroid Research (LINEAR) survey, and present our analysis results.
In Section \ref{sec:aspects} we discuss various aspects of a periodicity search through simulations.
Last, we provide concluding remarks and outline directions for future work in Section \ref{sec:conclu}.

\section{Bayesian inference and model selection of time series}
\label{sec:method}
In this section, we describe the framework of Bayesian inference and model selection for the analysis of time-series data.
We focus on the case of searching for periodicity in quasar light curves---quasar brightness measurements as a function of time.
Assuming stationary Gaussian noise, the likelihood function for a quasar light curve is
\begin{eqnarray}
\label{eq:likel}
\mathcal{L}( \mathbf{d} | \vec{\theta}_{n}, \vec{\theta}_{s} ,m) &= & \frac{1}{\sqrt{(2\pi)^{N} |\mathbf{C}|}} \\ && \exp\left[-\frac{1}{2} (\mathbf{d}-\mathbf{m}-\mathbf{s})^T \mathbf{C}^{-1} (\mathbf{d}-\mathbf{m}-\mathbf{s})\right]  \nonumber \, ,
\end{eqnarray}
where $\mathbf{d}$ is the time-series light curve data with length $N$, $\vec{\theta}_{n}$ and $\vec{\theta}_{s}$ include the noise and signal parameters, respectively. In this work we use measurements of optical magnitudes (i.e., the logarithmic light curve). The data $\mathbf{d}$ is modeled as
\begin{equation}
\label{eq:data}
    \mathbf{d}=\mathbf{n}+\mathbf{m}+\mathbf{s}\, ,
\end{equation}
where $\mathbf{n}$ is the noise vector, which contains measurement uncertainties and additional intrinsic stochastic quasar variability; $\mathbf{m}$ is a constant vector with identical entries of $m$, and $m$ accounts for the mean magnitude and any constant offset (e.g., constant level of contamination due to host galaxy light); the signal vector $\mathbf{s}$ is given by
\begin{equation}
\mathbf{s}(t)=A\sin(2\pi f_{0}t+\phi)\, .
\label{eq:signalt}
\end{equation}
The signal parameters are $\vec{\theta}_{s}=\{A,\phi,f_{0}\}$, where the signal frequency $f_0$ is related to the period by $f_{0}=1/T_{0}$.
In Equation (\ref{eq:likel}), $\mathbf{C}_{ij}=\langle \mathbf{n}_{i}\mathbf{n}_{j} \rangle$ is the noise covariance matrix, which contains two components, $\mathbf{C}=\mathbf{C}^{w}+\mathbf{C}^{r}$, where $\mathbf{C}^{w}$ is a diagonal matrix that represents white noise, and $\mathbf{C}^{r}$ accounts for the stochastic quasar variability, usually termed ``red noise."

The white noise matrix takes the following form
\begin{equation}
\mathbf{C}_{ij}^{w} = (\nu \sigma_{i})^2 \delta_{ij}\, ,
\label{eq:Cwhite}
\end{equation}
where $\sigma_{i}$ is the measurement uncertainty for the $i$th observation, $\nu$ is a scale factor used to quantify the over/under-estimation of measurement uncertainties\footnote{This is equivalent to the EFAC (Error FACtor) parameter used in pulsar timing data analysis. An additional EQUAD (Error added in QUADrature) parameter could be introduced to account for potential systematic errors that are not included in measurement uncertainties.}, $\delta_{ij}$ is the Kronecker delta function.

In the DRW model \citep{Kelly09DRW}, the covariance function is an exponential function. The red-noise covariance matrix is
\begin{equation}
\mathbf{C}_{ij}^{r} = \frac{1}{2}\hat{\sigma}^2\tau_{0}\exp\left(-\frac{\tau_{ij}}{\tau_0}\right)\, ,
\label{eq:Cdrw}
\end{equation}
where $\hat{\sigma}^2$ is the intrinsic variance between observations on short timescales ($\sim 1$ day), and $\tau_0$ is the damping timescale, and $\tau_{ij}\equiv|t_{i}-t_{j}|$.
The noise power spectral density of the DRW model is
\begin{equation}
    P(f)=\frac{2\hat{\sigma}^2\tau_{0}^2}{1+(2\pi\tau_{0} f)^2}\, .
\label{eq:psd_drw}
\end{equation}

To account for the possibility that the quasar red noise deviates from the DRW model \citep[e.g.,][]{Zu13_drw,Guo17_drw}, we also consider the following form of noise covariance matrix:
\begin{equation}
\mathbf{C}_{ij}^{r} = \frac{1}{2}\hat{\sigma}^2\tau_{0}\exp\left[-\left(\frac{\tau_{ij}}{\tau_0}\right)^{\gamma}\right]\, .
\label{eq:CVM_pe}
\end{equation}
This is also called the stretched exponential function\footnote{In the context of describing relaxation in disordered systems, it is called the Kohlrausch-Williams-Watts function.}.
Note that the special case of $\gamma=1$ corresponds to the DRW model, $\gamma=0$ means white noise, and $\gamma=2$ reduces to the Gaussian function.

We further extend our red-noise covariance matrix to account for the QPO phenomenon
\begin{equation}
\mathbf{C}_{ij}^{r} = \frac{1}{2}\hat{\sigma}^2\tau_{0}\exp\left[-\left(\frac{\tau_{ij}}{\tau_0}\right)^{\gamma}\right]\cos\left(\frac{2\pi \tau_{ij}}{T_{q}}\right)\, ,
\label{eq:CVM_qpo}
\end{equation}
where $T_{q}$ is the period of the QPO.
In the case of $\gamma=1$, the power spectral density of the above covariance function becomes
\begin{equation}
    P(f)=\frac{\hat{\sigma}^2\tau_{0}^2}{1+4\pi^{2}\tau_{0}^{2}(f-f_{q})^2}+\frac{\hat{\sigma}^2\tau_{0}^2}{1+4\pi^{2}\tau_{0}^{2}(f+f_{q})^2}\, ,
\label{eq:psd_qpo}
\end{equation}
where $f_{q}=1/T_{q}$ is the QPO frequency.
Note that in the limit of $T_{q} \rightarrow \infty$, the above equation reduces to Equation (\ref{eq:psd_drw}).
In practice, this model is indistinguishable from the DRW model once $T_q$ is longer than the data span. The noise parameters are $\vec{\theta}_{n}=\{\nu, \hat{\sigma}, \tau_{0}, \gamma, T_{q}\}$.
We illustrate the red-noise models and discuss their phenomenology in detail in Appendix \ref{sec:app_psd_cov}.

We use Bayesian model selection to quantify the statistical significance of the presence of periodic signals in quasar light curves. We start with Bayes' theorem, which states that
\begin{equation}
P(\vec{\theta}|\mathbf{d},\mathcal{H})= \frac{\mathcal{L}(\mathbf{d}|\vec{\theta},\mathcal{H})P(\vec{\theta}|\mathcal{H})}{{\cal Z}(\mathbf{d}|\mathcal{H})}\, .
\end{equation}
Here $P(\vec{\theta}|\mathbf{d},\mathcal{H})$ is the posterior probability distribution function of parameters $\vec{\theta}$ given data $\mathbf{d}$ and hypothesis $\mathcal{H}$; $\mathcal{L}(\mathbf{d}|\vec{\theta},\mathcal{H})$ is the likelihood function given in Equation (\ref{eq:likel}), which describes the probability of observing data given the hypothesis $\mathcal{H}$ and parameters $\vec{\theta}$. 
Meanwhile, $P(\vec{\theta}|\mathcal{H})$ is the prior distribution of parameters $\vec{\theta}$ while ${\cal Z}(\mathbf{d}|\mathcal{H})$ is the Bayesian evidence for hypothesis $\mathcal{H}$
\begin{equation}
\label{eq:BayesZ}
{\cal Z}(\mathbf{d}|\mathcal{H}) = \int \text{d}\vec{\theta} \mathcal{L}(\mathbf{d}|\vec{\theta},\mathcal{H})P(\vec{\theta}|\mathcal{H})\, .
\end{equation}

\begin{figure*}[ht]
\begin{center}
  \includegraphics[width=0.9\textwidth]{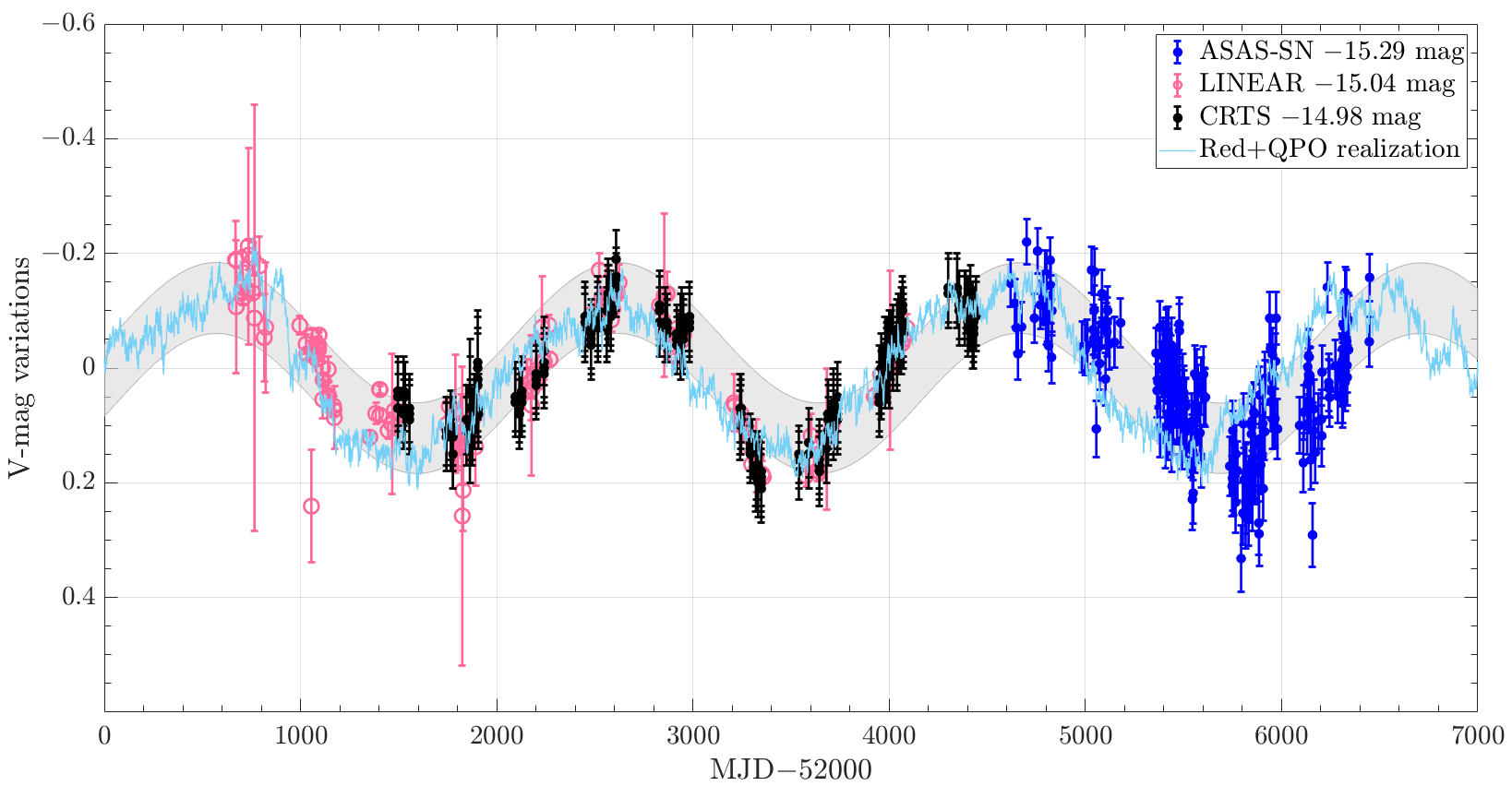}
  \caption{The light curve of PG1302$-$102 from LINEAR (pink), CRTS (black) and ASAS-SN (blue), with mean magnitudes subtracted. The grey band shows the 1-$\sigma$ uncertainty of the sinusoidal signal inferred from data. Also shown is a red-noise realization that contains a QPO.}
  \label{fig:PG1302data}
\end{center}
\end{figure*}

Given the observational data, we wish to compare two hypotheses: $\mathcal{H}_{n}$ = the data are consistent with only noise (i.e., $\mathbf{s}=0$), and $\mathcal{H}_{s}$ = there is a periodic signal present in the data. 
The ratio of posterior probability, called the \textit{odds ratio}, between hypotheses $\mathcal{H}_{s}$ and $\mathcal{H}_{n}$ is:
\begin{equation}
\mathcal{O}=\frac{P(\mathcal{H}_{s}|\mathbf{d})}{P(\mathcal{H}_{n}|\mathbf{d})} = \frac{\mathcal{Z}(\mathbf{d}|\mathcal{H}_{s})P(\mathcal{H}_{s})}{\mathcal{Z}(\mathbf{d}|\mathcal{H}_{n})P(\mathcal{H}_{n})}\, ,
\end{equation}
where $P(\mathcal{H}_{n})$ and $P(\mathcal{H}_{s})$ are the prior probability for hypotheses $\mathcal{H}_{n}$ and $\mathcal{H}_{s}$, respectively. Assuming equal prior probability for both hypotheses, Bayesian model selection is usually performed by computing the Bayes factor:
\begin{equation}
\mathcal{B}^{s}_{n} = \frac{\mathcal{Z}(\mathbf{d}|\mathcal{H}_{s})}{\mathcal{Z}(\mathbf{d}|\mathcal{H}_{n})}\, .
\end{equation}

In this work, we are concerned with the support for a sinusoidal signal quantified by $\mathcal{B}^{s}_{n}$.
We also wish to compute the support for the presence of QPO by comparing two models involving Equations (\ref{eq:CVM_qpo}) and (\ref{eq:CVM_pe}), and for the deviation from the DRW model, by comparing two models involving Equations (\ref{eq:CVM_pe}) and (\ref{eq:Cdrw}).
Following \cite{Raftery95}, the interpretation of Bayes factors is as follows. In a natural logrithmic scale, $0<\ln \mathcal{B}<1$ indicates that the support (for the hypothesis in the numerator) ``worth no more than a bare mention,'' $1<\ln \mathcal{B}<3$ implies positive support, $3<\ln \mathcal{B}<5$ implies strong support and $\ln \mathcal{B}>5$ implies very strong support.
These thresholds are somewhat arbitrary; $\ln \mathcal{B}=8$ is often adopted as the detection threshold in gravitational-wave astronomy \citep[e.g.,][]{ThraneTalbot19}.
Throughout this paper, the log evidence or log Bayes factor refer to the natural logarithm.

\begin{figure*}[ht]
\begin{center}
  \includegraphics[width=0.9\textwidth]{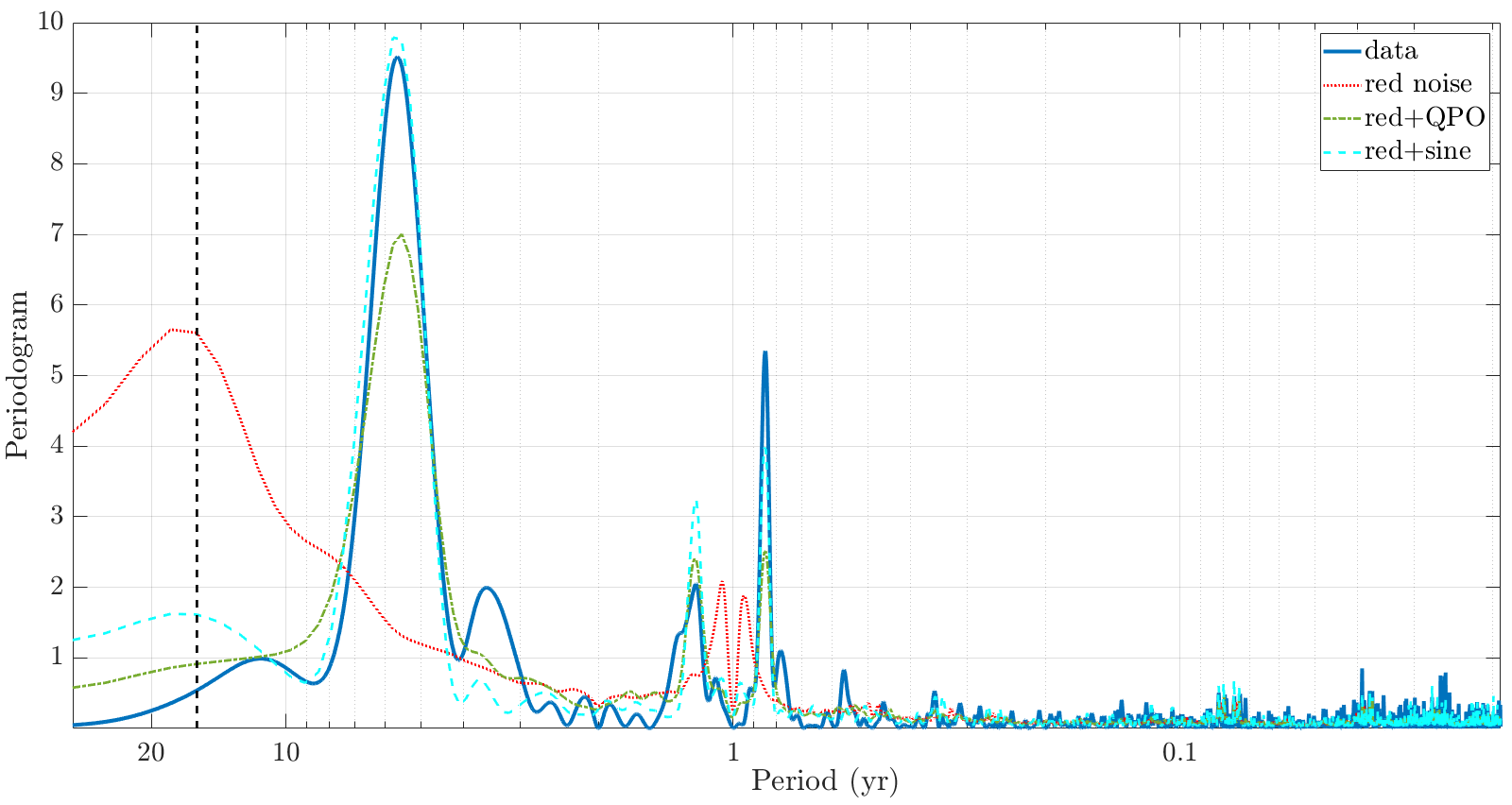}
  \caption{The Lomb-Scargle periodogram (in arbitrary unit) of PG1302$-$102 data. Also shown are the averaged periodogram from $10^3$ noise realizations (each generated with the same sampling as real data) for three models: red noise, red noise plus a QPO, and red noise plus a sinusoid. Model parameters are set at the median posterior probability values inferred from real data (see text). The vertical line marks the data span.}
  \label{fig:PG1302perido}
\end{center}
\end{figure*}

\section{Application to PG1302$-$102}
\label{sec:pg1302}

PG1302$-$102 is a nearby bright quasar with a median V-band magnitude of $15.0$ at a redshift of $z=0.2784$.
We apply our method to V-band magnitude measurements collected with CRTS \citep{CRTS09}, ASAS-SN \citep{ASAS-SN14,ASAS-SN19}, and LINEAR \citep{LINEAR11}.
The CRTS data are publicly available\footnote{\url{http://nesssi.cacr.caltech.edu/DataRelease/}} as part of the Catalina Surveys Data Release 2, including 290 phototmetric measurements taken between 6 May 2005 and 30 May 2013.
The ASAS-SN data are downloaded from its photometry database\footnote{\url{https://asas-sn.osu.edu/photometry}} as part of their Data Release 9, including 232 measurements acquired between 23 November 2013 and 27 November 2018.
The LINEAR data contain 626 measurements made between 23 January 2003 and 12 June 2012.
The median measurement uncertainties are 0.06, 0.05 and 0.01 mag for CRTS, ASAS-SN and LINEAR data, respectively.
The span of the combined data set is $\unit[15.84]{years}$.
To reduce the computational cost, we average LINEAR data with an interval of $\unit[1]{day}$, which reduces the number of data points to $138$.
The bin size is chosen such that any stochastic time-correlated variations over a timescale greater than $\unit[1]{day}$ are preserved.
The new uncertainty for binned data is computed as the standard deviation of original measurements inside the binning window.
The median uncertainty for averaged LINEAR data is $\unit[0.015]{mag}$.

Figure \ref{fig:PG1302data} shows the data used in our analysis: LINEAR in pink, CRTS in black and ASAS-SN in blue. The error bars indicate the 1-$\sigma$ measurement uncertainties.
We also show the 68\% credible region of the sinusoidal signal (grey shaded band) with a period of $\unit[5.66_{-0.10}^{+0.17}]{years}$ inferred from the data, and a red-noise realization (thin light blue line) that contains a QPO with a period of $\unit[5.56]{years}$ and might fit the data well.

Before we present details of our analysis results, we show in Figure \ref{fig:PG1302perido} the Lomb-Scargle periodogram of PG1302$-$102 data (solid blue line), along with the averaged periodogram taken from $10^3$ noise realizations for three models: red noise (dotted red line), red noise plus a QPO (dash-dotted green line), and red noise plus a sinusoid (dashed cyan line).
The vertical black dash line indicates the data span of $\unit[15.84]{years}$ for PG1302$-$102.
We simulate data using the same sampling in time and error bars as real data to obtain the average periodogram for three models.
The following median posterior probability model parameters as inferred from real data are adopted: 1) $\ln \hat{\sigma}^2=-2.63\, \text{mag}^{2}\, \text{yr}^{-1}$, $\ln (\tau_{0}/\textrm{yr})=2.05$, $\gamma=0.74$, for the red-noise hypothesis; 2) $T_{0}=\unit[5.56]{yr}$, $\ln \hat{\sigma}^2=-3.47\, \text{mag}^{2}\, \text{yr}^{-1}$, $\ln (\tau_{0}/\textrm{yr})=2.94$, $\gamma=0.52$, for the ``red noise + QPO" hypothesis; 3) $A=\unit[0.13]{mag}$, $T_{0}=\unit[5.66]{yr}$, $\phi=\unit[1.64]{rad}$, $\ln \hat{\sigma}^2=-2.87\, \text{mag}^{2}\, \text{yr}^{-1}$, $\ln (\tau_{0}/\textrm{yr})=0.95$, $\gamma=0.57$, for the ``red noise + sinusoid" hypothesis.
The purpose of Figure \ref{fig:PG1302perido} is to provide some insights into how different models might fit the data, instead of a proper accounting of the goodness of these models.
The periodogram of real data exhibits a strong peak at $\unit[5.6]{years}$, with secondary peaks at around 1 year.
As one can see, the major peak can be reproduced with both a QPO and sinusoidal model but not the pure red-noise model.
The secondary peaks at around 1 year are probably artefacts of the irregular sampling of the data, since simulated data that contain no signal at that period also result in apparent peaks there.

%It may appear at first that our prior $c$hoice for sinusoidal amplitude $A$ disfavours the noise hypothesis ($A=0$).
%However, we find the difference in the computed evidence for the signal hypothesis is negligible when an alternative parameterization is used: $A_{1}\cos(2\pi t/T_{0})+A_{2}\sin(2\pi t/T_{0})$, with $A_1$ and $A_2$ taking Gaussian priors centered at zero, as adopted in \citet{Vaughan16_false}.

\subsection{Results}

To implement the analysis outlined in Section \ref{sec:method} for PG1302$-$102 data, we employ the \texttt{Bilby} software package \citep{Bilby19}, which is a general and versatile Bayesian inference library developed primarily for gravitational-wave astronomy.
For the stochastic sampling of posteriors and evidence calculation, we use the dynamic nested sampling method developed by \citet{DynamicNested19}, which is available through the \texttt{Dynesty} package \citep{Dynesty}.
The uncertainty of the logarithmic model evidence ($\ln \mathcal{Z}$) computed by \texttt{Dynesty} is $\mathcal{O}(0.1)$.
By performing independent runs of the same sampling process, we verify that the returned values of $\ln \mathcal{Z}$ are consistent within the uncertainty, and that the posterior distributions have converged.

\begin{table}[!htb]
\centering
 \caption{Priors used in the analysis.}
  \label{tab:PriorRanges}
 \begin{tabular}{|l|c|}
  \hline
 Parameter & Prior description \\
  \hline
    \multicolumn{2}{|c|}{Same for prior $a$ \& $b$} \\
 \hline
  $A$ (mag) & Uniform, min=0, max=0.5 \\
  $\phi$ (radian) & Uniform, min=0, max=$2\pi$ \\
  $T_0$ (yr) & Uniform, min=0, max=10 \\
  $m$ (mag) & Uniform, min=14.5, max=15.5 \\
  $\nu$ & Uniform, min=0.1, max=2 \\
  $\gamma$ & Uniform, min=0, max=2\\
  \hline
  \multicolumn{2}{|c|}{prior $a$} \\
  \hline
  $\ln \hat{\sigma}^2$ (mag$^2$ yr$^{-1}$) & Normal, mean=$-4.0$, width=1.15 \\
  $\ln \tau_0$ (yr) & Normal, mean=$-0.6$, width=1.15 \\
  \hline
  \multicolumn{2}{|c|}{prior $b$} \\
  \hline
  $\ln \hat{\sigma}^2$ (mag$^2$ yr$^{-1}$) & Uniform, min=$-6$, max=0 \\
  $\ln \tau_0$ (yr) & Uniform, min=$-4$, max=4 \\
  \hline
  \multicolumn{2}{|c|}{prior $c$, same as prior $b$, with $m$ and $\nu$ fixed.} \\
 \hline
 \end{tabular}
\end{table}

Our priors are specified in Table \ref{tab:PriorRanges}.
We consider three cases denoted by prior $a$, $b$, and $c$.
The $a$ prior employs log-normal priors for the DRW model parameters $\hat{\sigma}^2$ and $\tau_0$, same as those used in \citet{Vaughan16_false}, which were based on earlier studies of quasar red noise under the DRW model \citep{Macleod10_drw,Kozlowski10_drw,Andrae13_drw}.
The $b$ prior employs a log-uniform prior for both $\hat{\sigma}^2$ and $\tau_0$.
Specifically, the prior edges on $\tau_0$ correspond to $\unit[6.7] {days}$ and $\unit[54.6] {years}$.
For the remaining parameters, we assign uniform priors.
There is one mean magnitude $m$ and one white-noise scaling factor $\nu$ for each data set (CRTS, ASAS-SN and LINEAR).
Prior $c$ is the same as prior $b$, except that we fix these six parameters: $m$ is fixed at the weighted mean magnitudes - $\unit[14.98] {mag}$, $\unit[15.29] {mag}$, and $\unit[15.04] {mag}$ for CRTS, ASAS-SN, and LINEAR, respectively; $\nu$ is fixed at 0.22, 1.15 and 1.0 for CRTS, ASAS-SN, and LINEAR, respectively, based on analysis of the data under the best-performing hypothesis as we discuss below.
We find the posteriors on these six parameters are well constrained and do not change with respect to the choice of signal/noise terms included in the hypothesis or the choice of prior for red noise parameters.

In Table \ref{tab:logZpriors}, we show the effect of the choice of priors on $\ln \mathcal{Z}$ for the red-noise hypothesis.
We find that prior $b$ results in slightly higher $\ln \mathcal{Z}$ than prior $a$.
We also note that prior $b$ leads to posterior distribution of $\tau_0$ that is in the low-probability tail of the log-normal prior.
Therefore, we adopt the less-informative log-uniform prior on red-noise parameters $\hat{\sigma}^2$ and $\tau_0$.
By fixing $m$ and $\nu$, the evidence $\ln \mathcal{Z}$ increases by 14.6 for the full combined data set (comparing prior $c$ against prior $b$).
This shows that the inclusion of these parameters is unnecessary, and we only present results based on prior $c$ in the remaining sections unless otherwise specified.

\begin{table}[!htb]
\centering
 \caption{The differences in log evidence ($\ln \mathcal{Z}$) of the red-noise hypothesis under three priors (with respect to prior $a$), for different combinations of data sets: C -- CRTS, L -- LINEAR, A -- ASAS-SN.}
  \label{tab:logZpriors}
 \begin{tabular}{|l|c|c|c|c|}
  \hline
  & C & C+L & C+A & C+L+A \\
  \hline
  prior $a$ & 0 & 0 & 0 & 0 \\
  prior $b$ & 0.7 & 1.2 & 1.1 & 1.2 \\
  prior $c$ & 5.9 & 11.8 & 10.1 & 15.8 \\
  \hline
 \end{tabular}
\end{table}

\begin{table}[!htb]
\centering
 \caption{The log Bayes factors $\ln \mathcal{B}$ between pairs of hypothesis, for different combinations of data sets. The first six rows are compared with the red-noise-only hypothesis.}
  \label{tab:logZlogB}
 \begin{tabular}{|l|c|c|c|c|}
  \hline
  & C & C+L & C+A & C+L+A \\
  \hline
  1. red & 0 & 0 & 0 & 0 \\
  \hline
  2. red+sine & 9.1 & 14.3 & 4.6 & 12.7\\
  \hline
  3. DRW & 1.3 & $-5.0$ & 1.6 & $-4.0$ \\
  \hline
  4. DRW+sine & 6.1 & 0.2 & 4.0 & 1.3 \\
  \hline
  5. red+QPO & 9.3 & 16.5 & 6.9 & 14.5 \\
  \hline
  6. red+QPO+sine & 9.2 & 17.2 & 6.2 & 14.1 \\
  \hline
  \hline
  $\ln \mathcal{B}_{4}^{2}$ & 3.0 & 14.1 & 0.6 &	11.4 \\
  \hline
 \end{tabular}
\end{table}

In comparison to \citet{Vaughan16_false}, we find that the DRW model is overwhelmingly favoured over the sinusoidal model, with $\ln \mathcal{B}=180$ ($\log_{10}\mathcal{B}=78$) using CRTS data alone (under prior $a$).
This is consistent with the result of $\log_{10}\mathcal{B}>60$ obtained in \citet{Vaughan16_false}.
However, it may be argued that the choice between pure red noise and pure sinusoid is a false dilemma as we expect a realistic signal to be characterized by both red noise \textit{and} a sinusoidal signal.
The question investigated here therefore is whether or not there is a sinusoidal signal \textit{on top of} red noise.
The corresponding physical picture is that fluctuations in the accretion disc of the primary black hole (or the circumbinary disc) produces red noise and the binary motion results in periodic variations.

In Table~\ref{tab:logZlogB}, we list the log Bayes factor $\ln \mathcal{B}$ for six hypotheses with respect to the red-noise only hypothesis (Eq. \ref{eq:CVM_pe}, denoted as ``red").
We examine the presence of a sinusoidal signal (denoted as ``sine") on top of additional two noise scenarios: the DRW model (Eq. \ref{eq:Cdrw}) and the red noise plus a QPO term (Eq. \ref{eq:CVM_qpo}, denoted as ``red+QPO").
Moreover, $\ln \mathcal{B}_{4}^{2}$ compares the ``red+sine" hypothesis to the ``DRW+sine" hypothesis, and indicates the level of support for deviation from the DRW red noise.

\begin{figure*}[ht]
\begin{minipage}[t]{0.5\textwidth}
\includegraphics[width=0.9\linewidth]{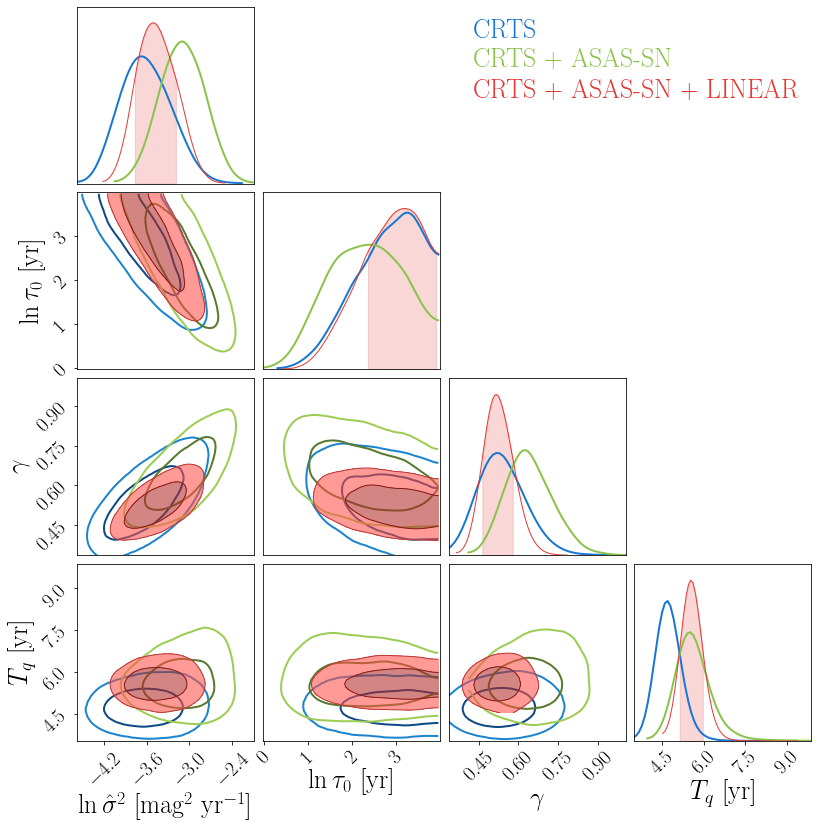}
\end{minipage}
\begin{minipage}[t]{0.5\textwidth}
\includegraphics[width=0.9\linewidth]{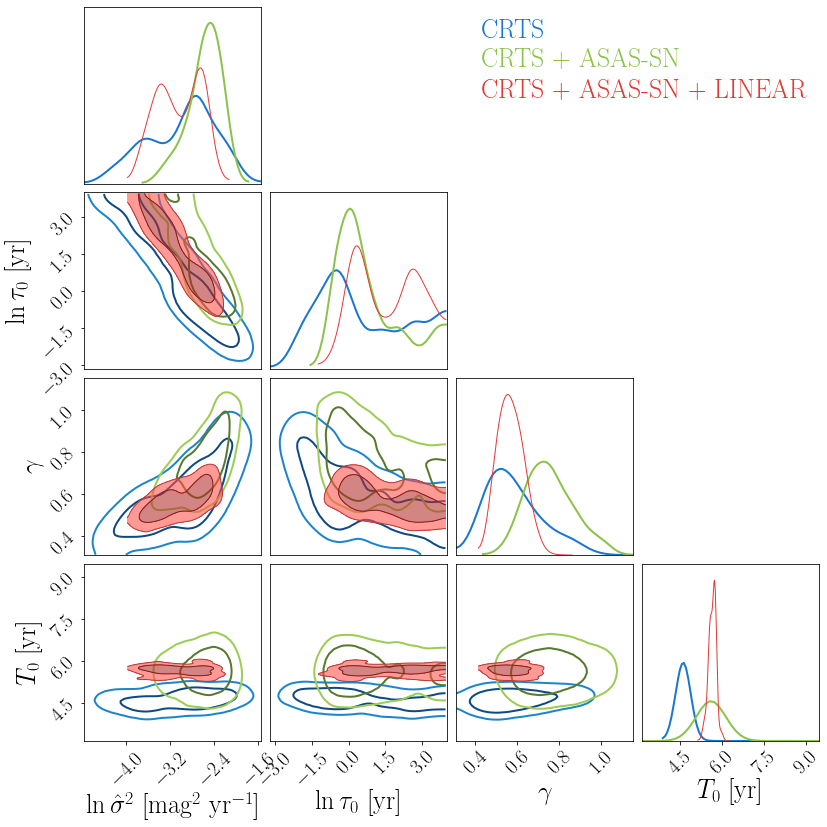}
\end{minipage}
\caption{Posterior distributions of model parameters inferred from different combinations of data sets for two hypotheses: the ``red+QPO" hypothesis (\textit{left}) and the ``red+sine" hypothesis (\textit{right}).}
\label{fig:2corners}
\end{figure*}

There are a few take-away messages from Table \ref{tab:logZlogB}. First, there is very strong evidence against the noise-only hypothesis.
Using the full data set (C+L+A), we find strong support for either a sinusoidal signal ($\ln \mathcal{B}=12.7$) or a QPO term ($\ln \mathcal{B}=14.5$), with the QPO slightly favoured with $\ln \mathcal{B}=1.8$.
Assuming equal prior probability for both signal hypotheses, the QPO hypothesis is favoured by an odds ratio of $6.0$.
While this is inconclusive, we discuss other hints toward the QPO later on.
Second, the DRW red-noise model is strongly disfavoured, with $\ln \mathcal{B}=11.4$ (a Bayes factor of $8.9\times 10^{4}$) using the full data set.
Third, there is a moderate reduction in $\ln \mathcal{B}$ associated with the inclusion of ASAS-SN data for almost all Bayes factors.
For the sinusoidal hypothesis, this is interpreted as evidence against the periodicity by \citet{LiuT18}.
We look into such a feature in more detail in Section \ref{sec:aspects}.
Last, there is no support for the presence of sinusoidal signal on top of the ``red+QPO" scenario.

The full posterior distribution of model parameters for the best performing hypothesis (``red+QPO", using prior $b$) is presented in Appendix \ref{sec:app_posterior}.
We find the CRTS measurement uncertainties are significantly overestimated by a factor of 5 ($\nu=0.20_{-0.01}^{+0.01}$), whereas those of ASAS-SN are slightly underestimated ($\nu=1.16_{-0.06}^{+0.07}$).
The unusual value of $\nu$ for CRTS data is consistent with that found in \citet{Vaughan16_false}, and can be verified by binning the original measurements inside a $\unit[1]{day}$ window and computing the dispersion.
For LINEAR data, we find $\nu=0.91_{-0.11}^{+0.12}$, which is consistent with $1$.

Figure \ref{fig:2corners} shows the posterior distributions for the ``red+QPO" (left panel) and the ``red+sine" hypothesis (right panel), for three combinations of data sets.
It can be seen that the use of the full data set generally results in tighter constraints on model parameters.
For the ``red+QPO" hypothesis, the exponential index $\gamma$ is measured to be $0.52_{-0.05}^{+0.06}$ using the full data set.
This rules out $\gamma=1$ (the DRW model) with more than $99.9\%$ credibility, consistent with the Bayes factor results in Table~\ref{tab:logZlogB}.
This measurement of $\gamma$, which is consistent with that for the ``red+sine" hypothesis, implies that the red noise in PG1302$-$102 features a power-law spectral density with a power-law index less than 2.
The red-noise time scale $\tau_{0}$ is long, $\gtrsim \unit[10]{yr}$, which is in the low-probability tail of the log-normal prior distribution (our prior $a$) as adopted in \citet{Vaughan16_false}.

Comparing two panels in Figure \ref{fig:2corners}, we find that, 1) posteriors of the ``red+QPO" hypothesis are better constrained; 2) there appear to be two posterior modes for ``red+sine" hypothesis for the red-noise amplitude $\hat{\sigma}^2$ and timescale $\tau_0$.
These two parameters are correlated, and one of the two modes -- large $\tau_0$ and small $\hat{\sigma}^2$ -- is consistent with the posterior distribution found under the ``red+QPO" hypothesis; 3) the posteriors for three combinations of data sets are consistent for the ``red+QPO" hypothesis, whereas those for the ``red+sine" hypothesis are unstable. 
Specifically, the credible region for the sinusoidal period $T_0$ is inconsistent between CRTS ($\unit[4.54_{-0.13}^{+0.16}$]{yr}) and the full data set ($\unit[5.66_{-0.10}^{+0.17}$]{yr}).
This comparison is an example of what is known as a posterior predictive check, in which one performs sanity tests to make sure the models are suitable for Bayesian inference.
The instability of posterior distributions with the inclusion of more data may suggest the ``red+sine" hypothesis does not fully describe the data.

\section{Discussion}
\label{sec:aspects}
\subsection{What does it mean that the significance of the periodicity in PG1302-102 goes down when we add more data?}

The growth in detection significance of a sinusoidal signal is not guaranteed in the presence of red noise.
A red-noise process produces long-term correlations in the data, meaning the noise component in new data is not independent from old data.
To demonstrate this effect, we inject a periodic signal into 10 random realizations of quasar red noise.
We choose exactly the same sampling and error bars as shown in Figure \ref{fig:PG1302data}.
Each of the 10 data sets contain three subsets, from LINEAR, CRTS and ASAS-SN.
We compute the Bayes factor, comparing the ``red+sine" hypothesis against the red-noise-only hypothesis, for different combinations of subsets.

We choose the following parameters, which are consistent with posterior estimates of PG1302$-$102: $A=0.12$ mag, $T_{0}=\unit[5.53]{yr}$, $\phi=\unit[5.0]{rad}$, $\ln \hat{\sigma}^2=-2.56\, \text{mag}^{2}\, \text{yr}^{-1}$, $\ln (\tau_{0}/\textrm{yr})=0.15$, $\gamma=0.62$, $\nu_{\text{CRTS}}=0.21$, $\nu_{\text{LINEAR}}=0.96$ and $\nu_{\text{ASAS-SN}}=1.15$.
The red noise realization is generated using $\mathbf{n}=\mathbf{L}\mathbf{r}$. 
Here $\mathbf{L}$ is a lower triangular matrix obtained from the Cholesky decomposition of the noise covariance matrix $\mathbf{C}=\mathbf{L}\mathbf{L}^{T}$ where $\mathbf{L}^{T}$ is the conjugate transpose of $\mathbf{L}$.
The $N\times 1$ vector of $\mathbf{r}$ contains $N$ independent random numbers that follow the standard normal distribution.

\begin{figure}[ht]
\begin{center}
  \includegraphics[width=0.46\textwidth]{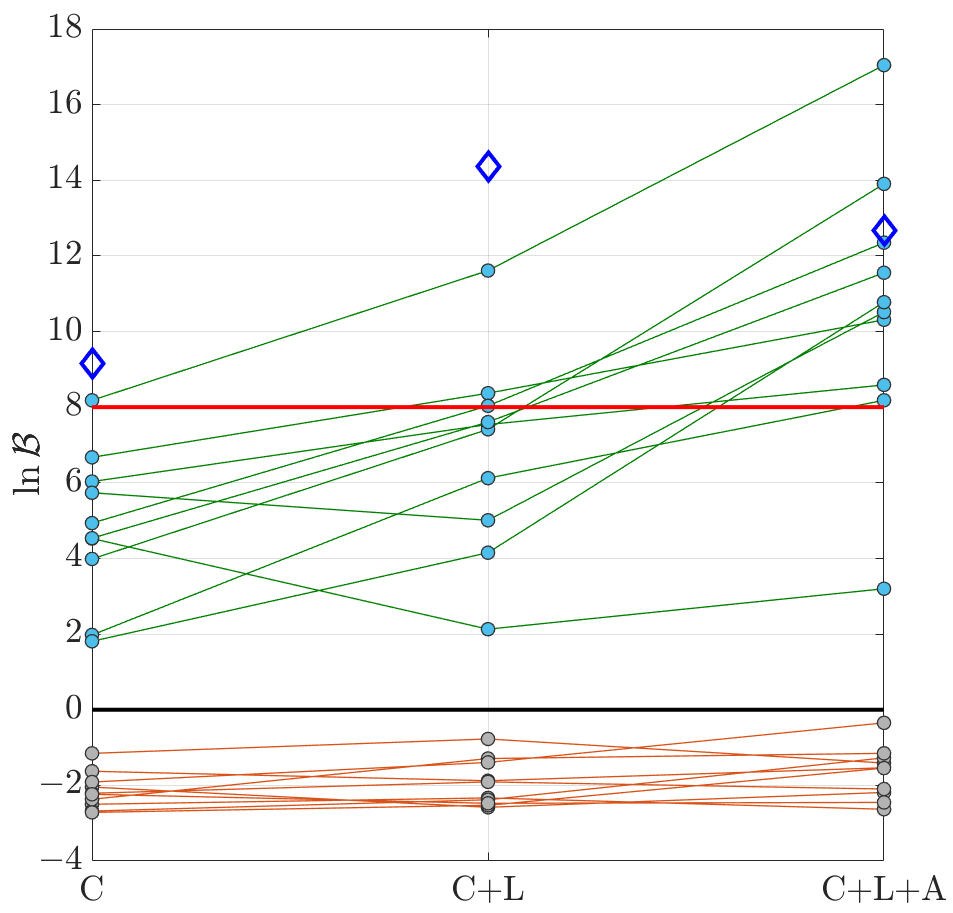}
  \caption{The log Bayes factor ($\ln \mathcal{B}$) that compares the red noise plus a sinusoidal signal hypothesis against the noise-only hypothesis for different combinations of data sets: C -- CRTS, A -- ASAS-SN, L -- LINEAR. Blue diamonds are for real data, whereas light blue (grey) dots are for simulated data that contain a sinusoidal signal (pure red noise). The red (black) horizontal line marks $\ln \mathcal{B}=8\,(0)$. \label{fig:PG1302BF}}
\end{center}
\end{figure}

Figure \ref{fig:PG1302BF} shows the Bayes factors from 10 simulated light curves of PG1302$-$102 as light blue dots with green lines.
Several features are noteworthy.
First, the general trend is that Bayes factors grow when we add LINEAR and ASAS-SN data to CRTS data.
Second, there are two noise realizations where the addition of LINEAR data onto CRTS data results in a reduced detection significance of periodicity.
We note that these two realizations have relatively low initial Bayes factors ($\ln \mathcal{B} \lesssim 6$).
Therefore, we conclude that 1) whether or not the periodicity significance grows with the inclusion of a certain set of additional data cannot be used as a simple criterion for true periodicity; 2) the Bayes factor is more likely to grow with time once the initial Bayes factor is high (e.g., $\ln \mathcal{B} \gtrsim 8$).
Third, the addition of ASAS-SN data generally leads to a greater improvement in detection significance than LINEAR.
As a sanity check, we also show the log Bayes factors for 10 noise-only data sets as grey dots in Figure \ref{fig:PG1302BF}, using the same noise parameters mentioned earlier.
In this case, the log Bayes factors are around $-2$, correctly disfavouring the presence of sinusoidal signals.
The addition of new data does not change the Bayes factor significantly.

\begin{figure}[ht]
\begin{center}
  \includegraphics[width=0.46\textwidth]{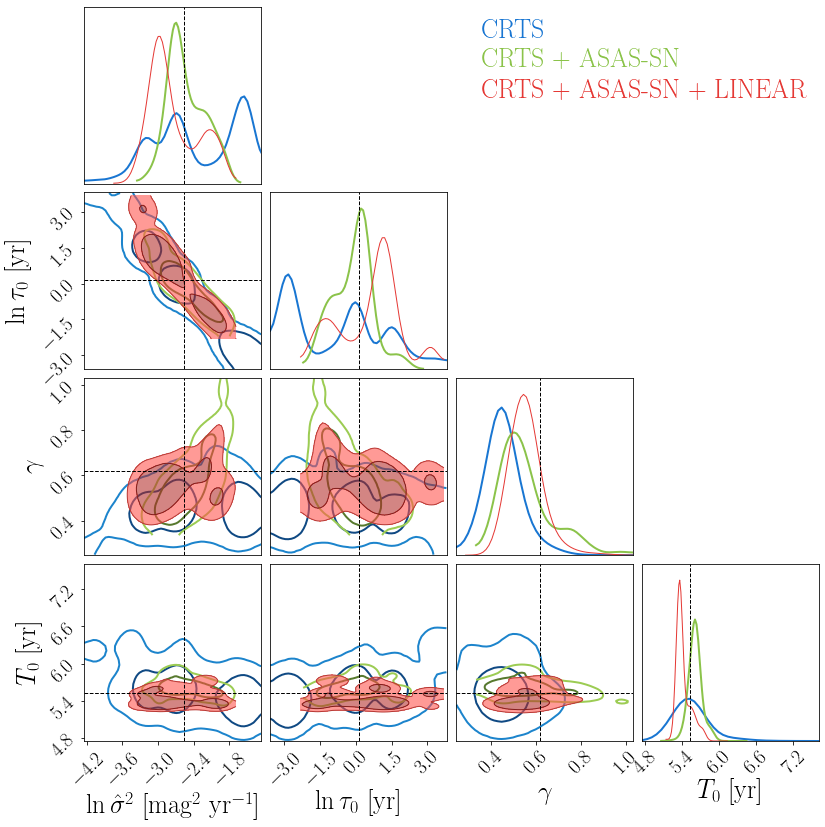}
  \caption{Same as the right panel of Figure \ref{fig:2corners} but for simulated data that contain an injected sinusoidal signal. Black dash lines mark the true values of parameters. \label{fig:rsg_corner_sim}}
\end{center}
\end{figure}

In Fig.~\ref{fig:PG1302BF}, we also show results from the analysis of real data as blue diamonds.
The evolution of $\ln \mathcal{B}$, particularly the reduction in $\ln \mathcal{B}$ associated with ASAS-SN data, is in drastic contrast with our simulations.
In Fig.~\ref{fig:rsg_corner_sim}, we plot the posterior distributions for one of the sinusoidal injections for different combinations of sub-data sets.
In comparison to the right panel of Figure \ref{fig:2corners}, which is the counterpart plot for real data, we find the posteriors are more stable.
Taken together, our simulations hint toward a problem with the sinusoidal signal hypothesis for PG1302$-$102.

Lastly, as a side note, the simulation study presented here highlights the challenge in the periodicity search in quasar light curves; namely, there is only one realisation of the red noise.
This situation is in contrast with the search for continuous gravitational waves (also essentially sinusoidal signals) using pulsar timing arrays.
Many millisecond pulsars in the timing array have been found to exhibit red noise for which the origin is largely unknown.
However, these red noise processes, if intrinsic to pulsars themselves, are not expected to correlate among different pulsars.
A false detection caused by red noise in one particular pulsar can be ruled out by cross-checking other pulsar data.
This sort of cross-check is not always possible when searching for binary black holes in quasar light curves.

\subsection{Can we distinguish a sinusoid from a QPO?}
Our analysis of PG1302$-$102 data reveals modest evidence for a QPO over a sinusoidal signal.
It is natural to ask: under what circumstances can we distinguish a true sinusoid from a QPO?
To find out, we compute the Bayes factor between the sinusoidal hypothesis and the QPO hypothesis for simulated data that contain a sinusoidal signal and some red noise.
We take the sampling and error bars of PG1302$-$102 data, and vary the signal period so that the data (spanning $\unit[15.84]{ yr}$) include 3, 6 and 9 wave cycles.
For Injection $a$, we choose the following parameters $A=\unit[0.13]{mag}$, $\phi=\unit[5.0]{rad}$, $\ln \hat{\sigma}^2=-2.56\, \text{mag}^{2}\, \text{yr}^{-1}$, $\ln (\tau_{0}/\textrm{yr})=0.15$, $\gamma=0.62$.
Injection $b$ is the same except that we increase the signal amplitude ($A=\unit[0.2]{mag}$) and reduce the red-noise amplitude ($\ln \hat{\sigma}^2=-4.56\, \text{mag}^{2}\, \text{yr}^{-1}$).

\begin{table}[!htb]
\centering
 \caption{The log Bayes factors that compare the ``red+sine" hypothesis against the red-noise-only hypothesis ($\ln \mathcal{B}_{\text{red}}^{\text{sine}}$), or the ``red+QPO" hypothesis ($\ln \mathcal{B}_{\text{QPO}}^{\text{sine}}$), for two sinusoidal signal injections into red noise. Results are listed for three different numbers of signal cycles. Injection $b$ contains a higher-amplitude signal than Injection $a$.}
  \label{tab:logB_sim}
 \begin{tabular}{|l|c|c|c|c|}
  \hline
  \multicolumn{2}{|c|}{Number of signal cycles} & 3 & 6 & 9 \\
  \hline
  \multirow{2}{*}{Injection $a$} & $\ln \mathcal{B}_{\text{red}}^{\text{sine}}$ & 15.0 & 25.5 & 36.4 \\
   & $\ln \mathcal{B}_{\text{QPO}}^{\text{sine}}$ & 0.1 & $-$2.4 & $-$2.1 \\
   \hline
  \multirow{2}{*}{Injection $b$} & $\ln \mathcal{B}_{\text{red}}^{\text{sine}}$ & 40.6 & 68.9 & 101.3 \\
   & $\ln \mathcal{B}_{\text{QPO}}^{\text{sine}}$ & 4.6 & 7.5 & 13.3 \\
  \hline
 \end{tabular}
\end{table}

\begin{figure}[ht]
\begin{center}
  \includegraphics[width=0.46\textwidth]{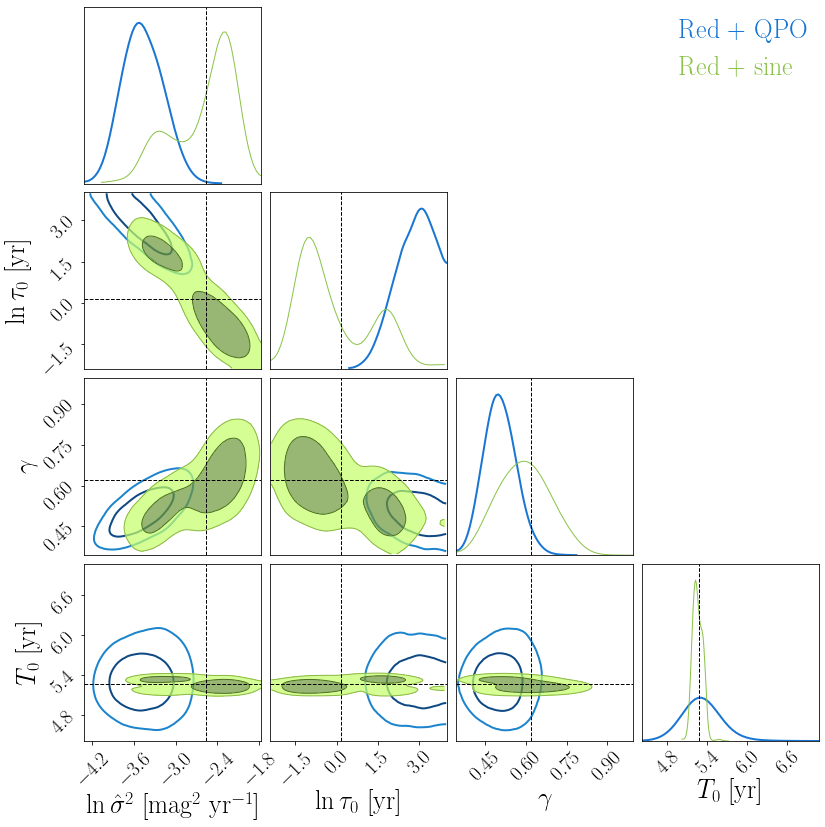}
  \caption{Posterior distributions of the ``red+QPO" and ``red+sine" hypotheses for a simulated data set that contains a sinusoidal signal on top of red noise (Injection $a$, the 3-cycle case in Table \ref{tab:logB_sim}). Black dash lines mark true values of parameters.}
  \label{fig:Sim_rsg_rgq}
\end{center}
\end{figure}

Table \ref{tab:logB_sim} lists the $\text{log}$ Bayes factors for both injections.
One can see that the Bayes factor that supports the presence of a sinusoidal signal increases with the number of signal cycles, as expected.
However, for Injection $a$, the sinusoid is indistinguishable from a QPO even after 9 wave cycles.
In fact, the Bayes factor slightly favours the QPO hypothesis for the 6-cycle and 9-cycle cases.
For the stronger Injection $b$, it becomes possible to tell the signal is a sinusoid instead of a QPO, and the corresponding Bayes factor grows with the number of wave cycles.

\begin{figure*}[ht]
\begin{center}
  \includegraphics[width=0.9\textwidth]{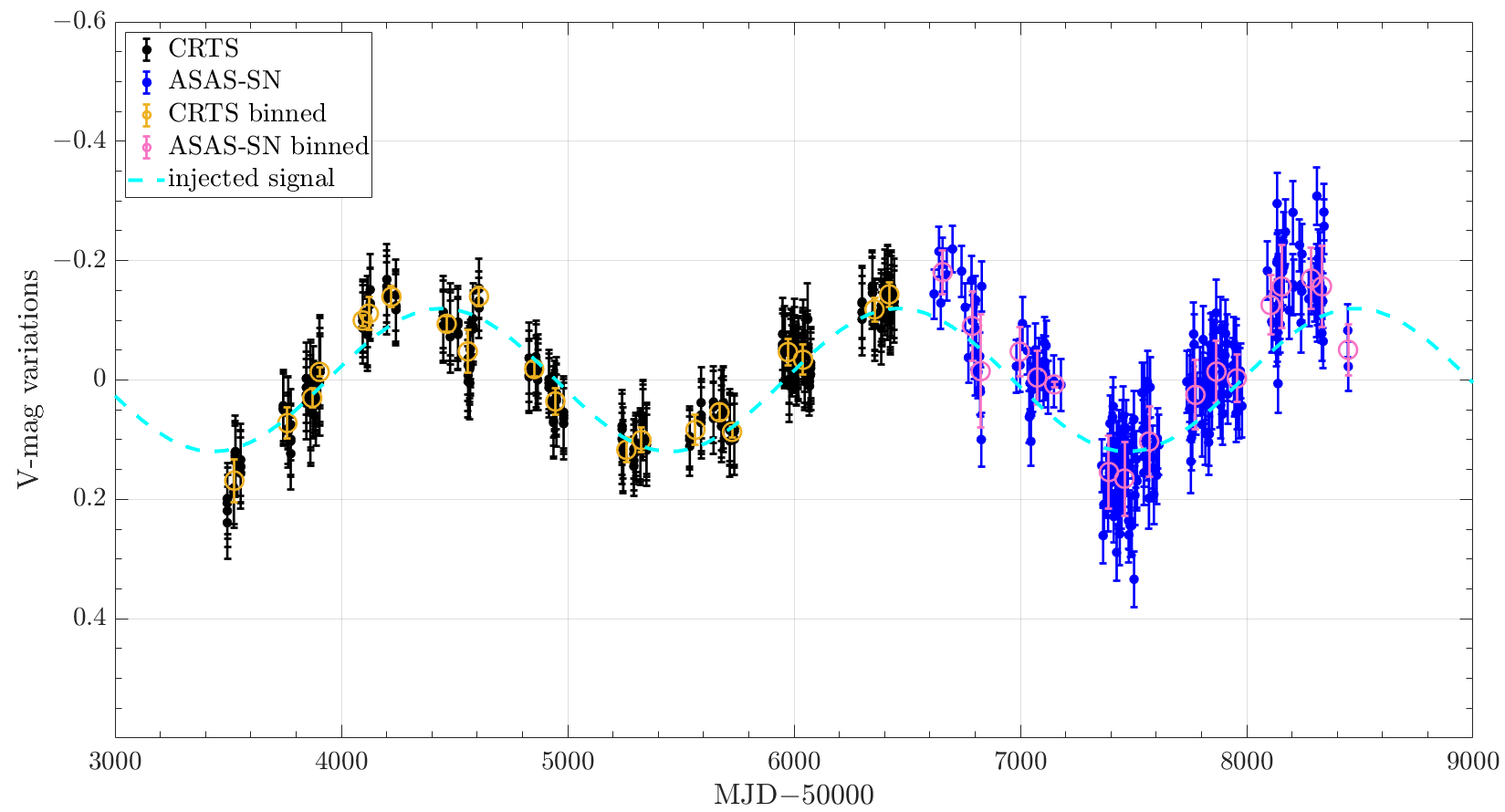}
  \caption{A simulated light curve of PG1302$-$102 that includes an injected sinusoidal signal. Mean magnitudes have been subtracted. Filled circles are original data, whereas open circles are binned data with an interval of 100 days.}
  \label{fig:PG1302data_sim}
\end{center}
\end{figure*}

Figure \ref{fig:Sim_rsg_rgq} shows the posterior distributions for the 3-cycle case of Injection $a$.
Whereas the injection parameters are correctly recovered for the true ``red+sine" hypothesis, red-noise parameters are mistaken by the (incorrect) ``red+QPO" hypothesis.
Noting that the quality factor of the QPO is given by $\pi \tau_{0}/T_{q}$, it is unsurprising that the QPO hypothesis results in posteriors that peak at large $\tau_0$ and small red-noise amplitude $\hat{\sigma}$.

\subsection{How does binning affect the detection significance and parameter estimation?}

It is common that some sort of averaging or binning is applied when analysing time-series data.
Information is inevitably lost in this process.
Here we demonstrate the potentially adverse effect of binning on the periodicity detection significance and parameter estimation precision with an example.
We choose the simulated data set that gives rise to the highest log Bayes factor in Figure \ref{fig:PG1302BF}.
Because the LINEAR data used in this work have already gone through an averaging process with an interval of $\unit[1]{day}$, we only use simulated data for CRTS and ASAS-SN.
Figure \ref{fig:PG1302data_sim} shows the simulated data, their binned versions with an interval of $\unit[100]{days}$, and the injected sinusoidal signal.

We compute two Bayes factors: a) the ``red+sine" hypothesis against the red-noise-only hypothesis, and b) the ``red+sine" hypothesis against the ``DRW+sine" hypothesis.
Cases a) and b) indicate the level of support for periodicity and for deviation from the DRW model, respectively.
The log Bayes factors are $13.0$ and $6.3$ for case a) and b), respectively, using the original data.
These Bayes factors become $7.7$ and $-0.1$ for case a) and b), respectively, using the binned data.
Therefore, the binning process not only reduces our ability to detect a periodicity but also to identify deviation from the DRW red-noise model (noting the high-frequency red-noise fluctuations illustrated in Figure \ref{fig:PG1302data}).

Posterior distributions for this injection study are shown in Appendix \ref{sec:app_posterior_sim} in Figures \ref{fig:PG1302_sim_CA} (unbinned data) and \ref{fig:PG1302_sim_CAbin} (binned data).
The binning process generally broadens the posterior distributions of both signal and noise parameters.
Note, in particular, that $\gamma$ posteriors for binned data are uninformative, which agrees with the Bayes factor ($\ln \mathcal{B}=0$) that shows the two red-noise models are indistinguishable.
A multimodal feature in the 2-D posterior of $T_{0}-\phi$ can be seen in Figure \ref{fig:PG1302_sim_CA}.
This can be attributed to the fact that we are estimating red noise and the periodic signal parameters simultaneously, and the data cover only a few ($\sim 3$) signal cycles.

\subsection{On the implementation of our method to a large sample of quasar light curves}
The computational cost in our analysis is dominated by the computation of the inverse and determinant of the $N \times N$ noise covariance matrix associated with the likelihood evaluation, which scales as $N^3$.
For the entire data set of PG1302$-$102 analysed in this work, $N=660$.
A single likelihood evaluation takes on the order of $\unit[0.1]{seconds}$ and the full parameter estimation and evidence calculation for such a data set take about $\unit[24]{hours}$ on a single modern CPU core.
The computational cost is the reason why we choose to average LINEAR data in an interval of $\unit[1]{day}$, which reduce the total number of data points from $1148$ to $660$.
Furthermore, there are a large number of quasars for which long-term photometric measurements are available, for example $10^5$ quasars from the CRTS as processed in \citet{Graham15}.
Therefore, further speed-up of our method is highly desirable.

This is a well-known problem in the analysis of astronomical time series.
In pulsar timing arrays, there are normally hundreds to thousands of times of arrival measurements for each pulsar and the problem involves correlation analysis with data from dozens of pulsars.
Various acceleration/approximation techniques have been proposed to enable full Bayesian analysis of pulsar timing array data.
One popular method is to approximate the red noise as the sum of $k$ Fourier components, and thus its covariance matrix is given by $\mathbf{C}=\mathbf{F} \mathbf{\Phi} \mathbf{F}^{T}$ where $\mathbf{F}$ is $N\times k$ (with $k \ll N$) and $\mathbf{\Phi}$ is a $k \times k$ diagonal matrix.
This turns the computationally heavy inversion of $\mathbf{C}$ into the lower-rank diagonal matrix inversion $\mathbf{\Phi}^{-1}$ through the Woodbury matrix lemma \citep[e.g.,][]{vanHaasVallisneri15,Lentati13}.
Elsewhere, \citet{Foreman17} proposed a fast Gaussian-process model that makes use of the feature where the covariance function can be expressed as a mixture of complex exponentials.
Implemented as the \texttt{Celerite} package, its computational cost scales linearly with the size of the data set; the same scaling for computational cost is shared by the CARMA method \citep{KellyCARMA14}.
All of these methods may prove useful for the analysis of light curves for a large number of quasars within the Bayesian framework developed here.

\section{Conclusions}
\label{sec:conclu}

Sub-parsec supermassive binary black holes are crucial in our understanding of galaxy evolution.
They are also the primary sources of nanohertz gravitational waves highly anticipated by international pulsar timing array campaigns.
While nearly impossible to resolve through direct imaging, such close binaries are expected to produce periodic variations in light curves of active galactic nuclei.
New candidates of periodicity are reported on a nearly monthly basis.

In this work, we propose a fully Bayesian method for the identification of periodicity in astronomical time-series that exhibit red noise.
We apply this method to one of the most promising periodicity candidates, PG1302$-$102, using data from CRTS, ASAS-SN and LINEAR surveys.
Our main findings are:
\begin{enumerate}
    \item There is very strong support for the presence of either a sinusoidal signal ($\ln \mathcal{B}=12.7$) or a QPO ($\ln \mathcal{B}=14.5$) at a period of $\unit[5.6]{yr}$. The data slightly favours the QPO hypothesis with an odds ratio of 6.
    \item The inclusion of ASAS-SN data reduces the log Bayes factor that supports the presence of a sinusoidal signal by 1.6. This, combined with the fact that the posterior distribution of sinusoidal period is unstable when we combine new data with CRTS, provides further evidence against the sinusoidal hypothesis.
    \item There is also conclusive evidence for deviation from the DRW red-noise model, with $\ln \mathcal{B}=11.4$. The noise power spectral density is shallower than a power law with an index of 2, and the red-noise damping timescale is long $\unit[\gtrsim 10]{yr}$.
\end{enumerate}
We perform simulations of sinusoidal signals and red noise and demonstrate the following:
\begin{enumerate}
    \item The growth of the periodicity significance with additional data is not guaranteed because of the stochastic fluctuations of red noise.
    For data sets like LINEAR, CRTS and ASAS-SN, the log Bayes factor is expected to grow once
    $\ln \mathcal{B} \gtrsim 8$ is achieved with initial data.
    \item Our method is capable of distinguishing a sinusoidal signal from a QPO.
    However, this is only possible for strong signals or a large number of wave cycles. A strong prior constraint on the red noise would also help.
    \item The use of data binning can reduce our ability to detect periodicity or deviation from the DRW model, because the binning throws away high-frequency information that helps estimate the spectral slope of the red noise (see Fig.~\ref{fig:PSD_cov_gamma}).
\end{enumerate}

We show that the data of PG1302$-$102 favours the presence of a quasi-periodicity at around $\unit[5.6]{years}$ against the noise hypothesis with a Bayes factor of $2\times 10^{6}$.
However, given that PG1302$-$102 was selected as the most periodic candidate out of $2\times 10^5$ quasars, it cannot be ruled out that such a quasi-periodicity is caused by pure red noise.
Assuming that it is indeed a QPO and the central black hole is $10^{8.5} M_{\odot}$ \citep{PG1302Nature}, we infer a period corresponding to a Keplerian orbit at 344 Schwarzschild radii ($\approx \unit[0.01]{pc}$).
We note that the QPO found in PG1302$-$102 roughly corresponds to the low-frequency QPOs seen in stellar-mass black hole X-ray binaries, although the QPO frequency of PG1302$-$102 lies one order of magnitude below the linear scaling relation between black hole mass and QPO frequency fitted to several candidates in \citet{Smith18_AGN-QPO}.
The physical origins of QPOs are poorly understood and therefore such a scaling relation might not exist if there are different mechanisms driving QPOs in small and big black holes.
Nevertheless, the QPO hypothesis of PG1302$-$102 can be tested with continued observations, and if the quasi-periodicity signature is confirmed, it may have significant implications for accretion physics of supermassive black holes.

Finally, our Bayesian framework can be adopted to establish unambiguous binary black hole detections with the following extensions:
\begin{enumerate}
    \item Apply this method to various physical models for the periodicity, such as periodic mass accretion rate of the binary \citep[e.g.,][]{Farris14_bbhaccretion} or jet precession of a single or binary supermassive black holes \citep[e.g.,][]{Abraham98_PreJet,Kudry11_PreJet,Britzen18_OJ187jet}, in addition to relativistic Doppler boosting;
    \item Use more sophisticated signal models that account for a cold-spot-induced perturbation in the accretion disc for PG1302$-$102 \citep{Kova19_pg1302perturb} or post-Newtonian orbital evolution for OJ 287 \citep{Dey18_oj287};
    \item Combine multi-wavelength observations \citep[e.g.,][]{Xin19_pg1302_swift}, and other signatures associated with a binary black hole, for example, Doppler velocity offsets in broad emission line profiles \citep{Eracle12_offsetLine,Bon12_NGC4151,LiYR16_NGC5548}, and flux deficits in the spectral energy distribution \citep{Gultekin_SMBBH_disk_cavity,Guo20_SED}. See \citet{ZhengZY16} for a binary black hole candidate for which different types of signatures are analyzed;
    \item Use astrophysically-motivated population priors.
    For example, the period distribution of binary black hole population is expected to be dominated by long periods, and the distribution of red-noise parameters can be obtained by applying our method to a large number of active galactic nuclei using hierarchical inference.
\end{enumerate}

\acknowledgements
We thank the anonymous referee for helpful comments that encouraged us to do a more thorough study on PG1302$-$102.
We thank Tingting Liu for sharing the PG1302 data used in \citet{LiuT18}, and Boris Goncharov and Yuri Levin for useful discussions.
This work is supported by ARC CE170100004
and ARC FT150100281.
The Catalina Sky Survey is funded by NASA under Grant No. NNG05GF22G issued through the Science Mission Directorate Near-Earth Objects Observations Program. The CRTS survey is supported by the U.S.~National Science Foundation under grants AST-0909182 and AST-1313422.
We use the \texttt{ChainConsumer} package \citep{Hinton2016} for some of the plots presented here.
Codes and data to reproduce major results of this work are publicly available at \url{https://github.com/ZhuXJ1/SuperBayes}.

\bibliography{ref}

\appendix

\section{Red-noise models}
\label{sec:app_psd_cov}

\begin{figure*}[!htb]
  \centering
  \includegraphics[width=0.9\textwidth]{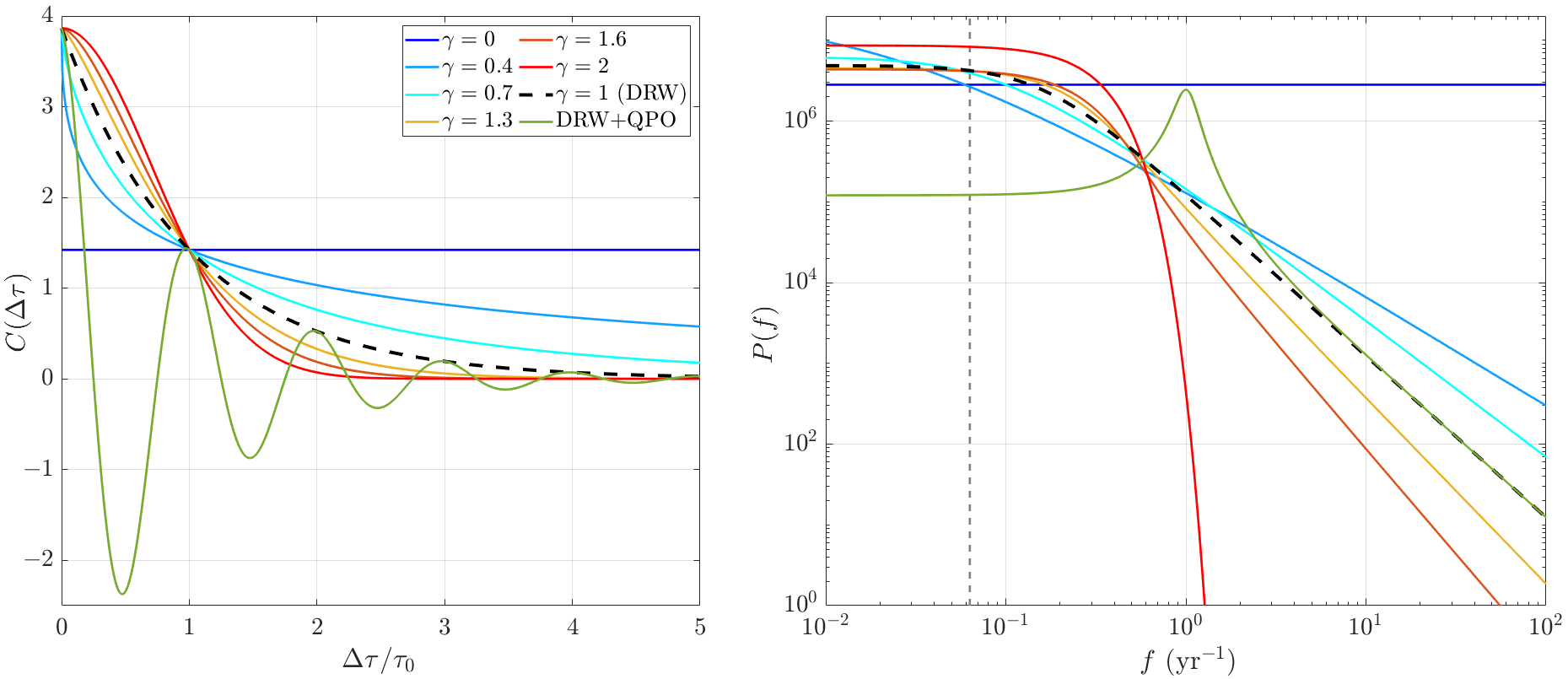}
  \caption{The covariance function $C(\Delta \tau)$ (\textit{left}) and power spectral density $P(f)$ (\textit{right}), both in arbitrary unit, for different values of $\gamma$ (see Equation \ref{eq:CVM_pe}), and for the presence of a QPO on top of the DRW red noise ($\gamma=1$, see Equation \ref{eq:CVM_qpo}). The grey vertical line on the right panel corresponds to the data span of $15.84$ yr for PG1302$-$102.}
  \label{fig:PSD_cov_gamma}
\end{figure*}

In Fig.~\ref{fig:PSD_cov_gamma}, we show the covariance function (left panel) and power spectral density (right panel) for the red-noise models considered in this work. Black dashed lines are for the DRW model, i.e., $\gamma=1$, and coloured lines are for different values of $\gamma$.
Additionally, we show the case of a QPO on top of the DRW red noise in green line.
The grey vertical line on the right panel indicates $1/T_{\text{obs}}$ where $T_{\text{obs}}=15.84$ yr is the data span of PG1302$-$102.
The time scale $\tau_0$ and the QPO period $T_q$ are both set to be 1 yr in these examples.

The prevalent DRW red-noise model features a spectral turnover at a frequency that corresponds to the damping timescale $\tau_0$, and a power-law spectral density with a power-law index of $2$ at high frequency ($\gtrsim 1/\tau_{0}$).
As can be seen in the right panel of Fig.~\ref{fig:PSD_cov_gamma}, our general red-noise model covers a broad range of power-law spectral shapes, from a power-law index of zero ($\gamma=0$) to extremely steep spectrum ($\gamma=2$).
In order to measure the spectral shape of quasar red noise, i.e., the $\gamma$ parameter, the high-frequency part is critical since that is where $\gamma$ has the largest influence.
In the case that $\tau_0 \gtrsim T_{\text{obs}}$, various models reduce to the power-law model.
The QPO model, on the other hand, is drastically different as it features a peak in the power spectral density. The quality factor of the QPO, defined as the ratio of peak frequency of the QPO
to its width, is determined by $\pi \tau_{0}/T_{q}$.

We note that the DRW model and the QPO prescription adopted here are special cases of the flexible CARMA (continuous-time autoregressive moving average) models. \citet{KellyCARMA14} demonstrated that CARMA models are useful in identifying new features in the power spectral density of astronomical time-series data.
Since the power spectral density of CARMA models can be expressed as a sum of Lorentzian functions, in a similar fashion to the \texttt{Celerite} model described in \citet{Foreman17}, it allows fast evaluation of the likelihood function which scales linearly with the number of data points.
This makes it particularly powerful for analyzing massive time-domain data from a large number of objects.
We leave the exploration of CARMA and \texttt{Celerite} models in our Bayesian inference framework for future work.

\section{Posterior distributions from analysis of real data}
\label{sec:app_posterior}

\begin{figure*}[!htb]
  \centering
  \includegraphics[width=\textwidth]{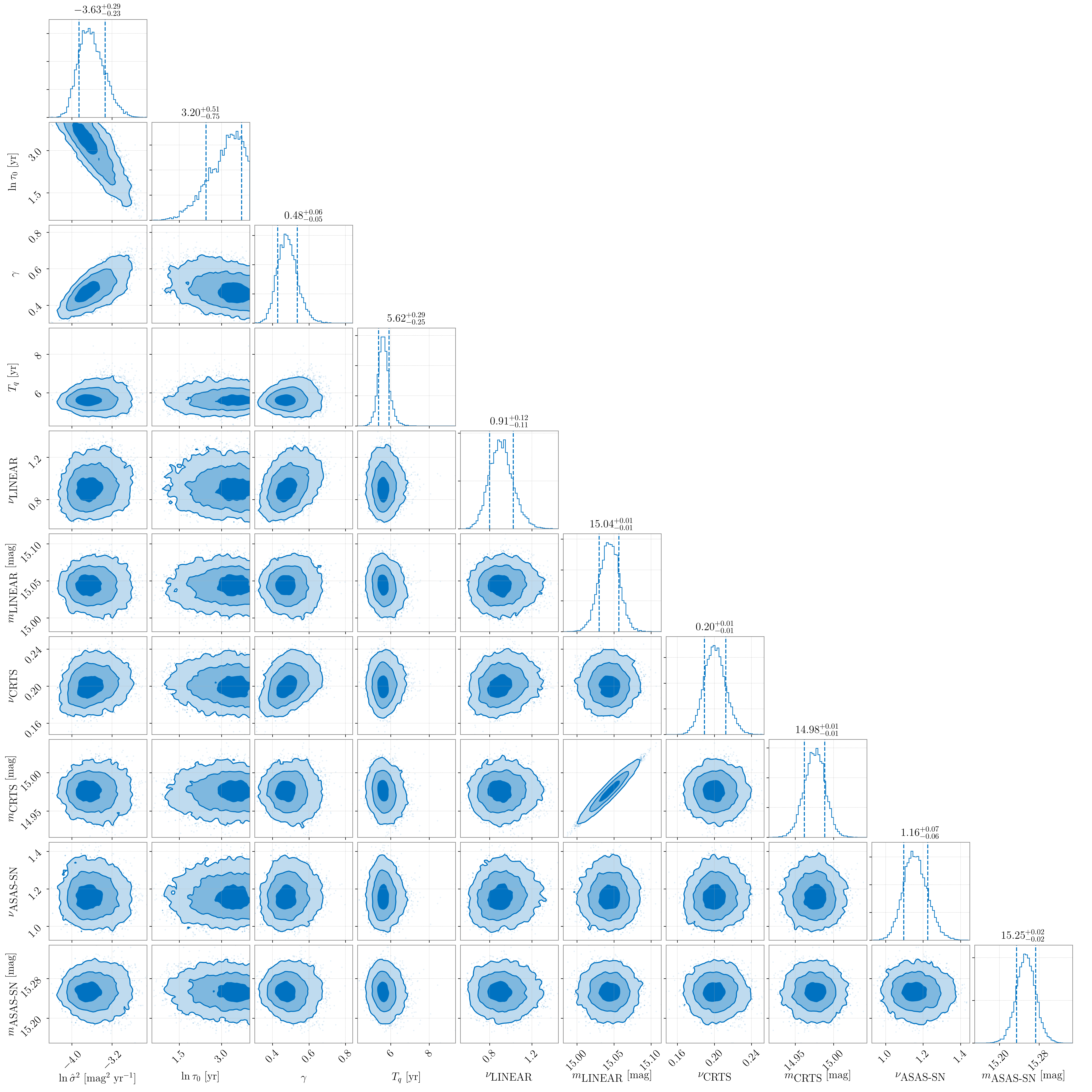}
  \caption{Posterior distributions of parameters for the red noise + QPO hypothesis for the light curve of PG1302$-$102.}
  \label{fig:PG1302red}
\end{figure*}

Here we present full posterior distributions from the analysis of the combined data set (Fig.~\ref{fig:PG1302data}) from CRTS, LINEAR and ASAS-SN for PG1302$-$102.
Fig. \ref{fig:PG1302red} shows the distributions for the red-noise plus a QPO hypothesis, which is the favoured hypothesis given the data.
Note that the mean offset parameter $m$ for LINEAR data and CRTS data are highly covariant because there is a significant overlap in time for the two.

\section{Posterior distributions from analysis of simulated data}
\label{sec:app_posterior_sim}

\begin{figure*}[!htb]
  \centering
  \includegraphics[width=\textwidth]{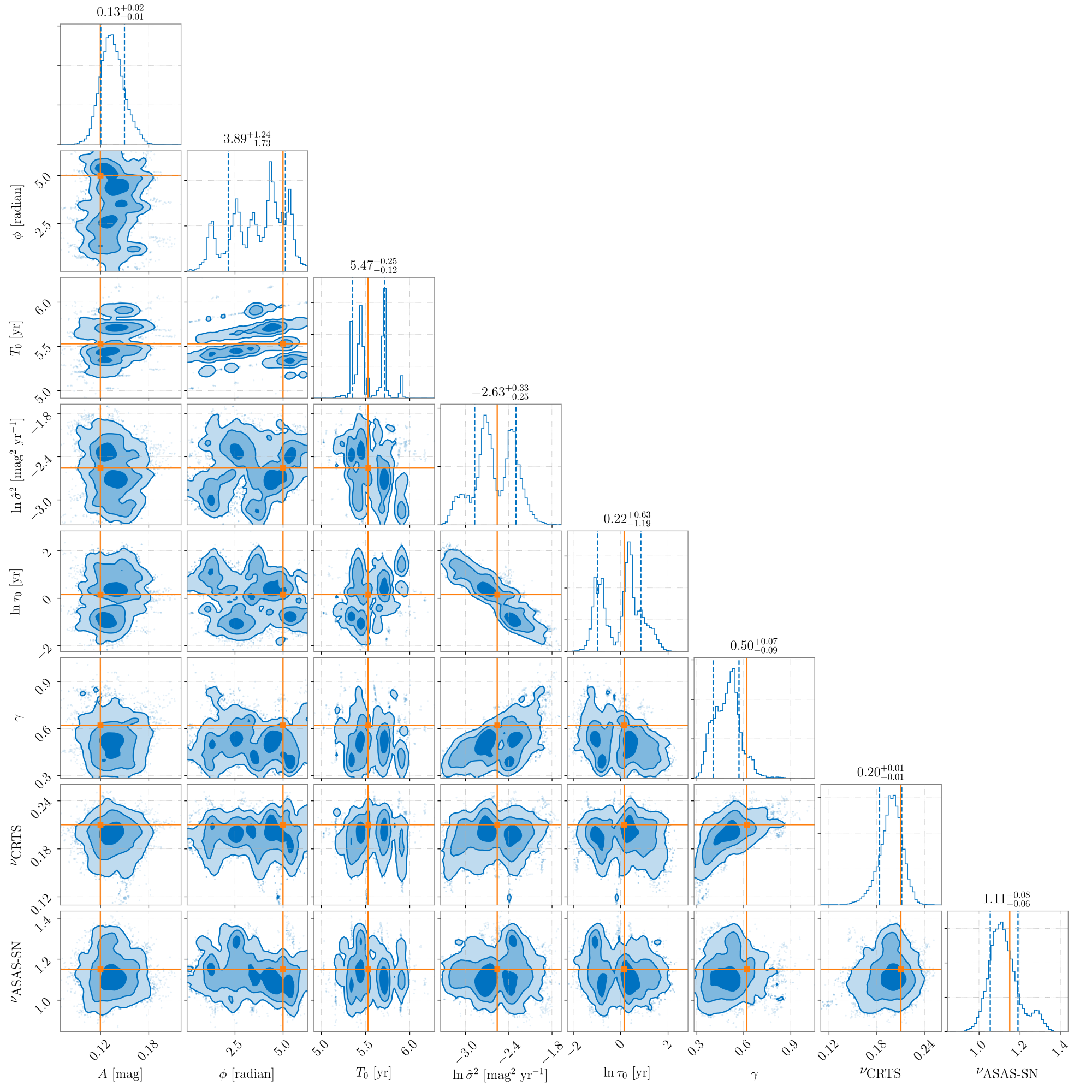}
  \caption{Posterior distributions of parameters for the sinusoidal signal plus red noise hypothesis for a simulated data set of PG1302$-$102 (shown in Figure \ref{fig:PG1302data_sim}), which included an injected signal. Orange lines mark the injection values.}
  \label{fig:PG1302_sim_CA}
\end{figure*}

\begin{figure*}[!htb]
  \centering
  \includegraphics[width=\textwidth]{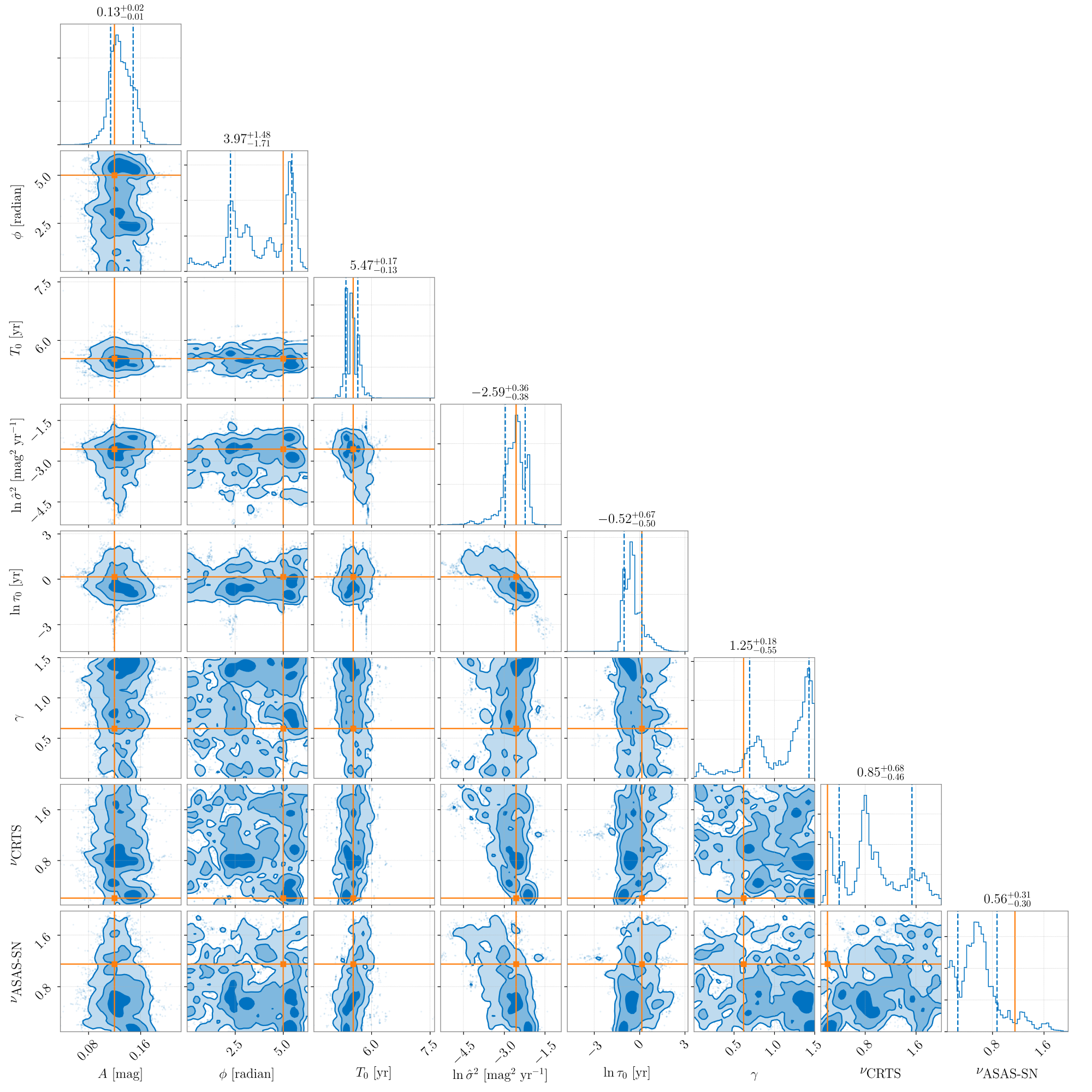}
  \caption{Same as Figure \ref{fig:PG1302_sim_CA} but for the binned data with an interval of 100 days.}
  \label{fig:PG1302_sim_CAbin}
\end{figure*}

Here we present full posterior distributions from the analysis of a simulated data set (Fig. \ref{fig:PG1302data_sim}) that includes an injected sinusoidal signal.
Fig. \ref{fig:PG1302_sim_CA} shows the distributions for original data, whereas Fig. \ref{fig:PG1302_sim_CAbin} shows results for the binned data with an interval of 100 days.

\end{document}